\documentstyle[12pt]{article}

\begin{document}
\def\bib#1{[{\ref{#1}}]}
\def\at{\tilde{a}}

\begin{titlepage}
         \title{ Maximal Acceleration Limits on the  \\
                 Mass of the Higgs Boson}

\author{G. Lambiase$^{a,}$\thanks{E-mail:lambiase@vaxsa.csied.unisa.it},
G.Papini$^{b,}$\thanks{E-mail:papini@cas.uregina.ca}  
and G. Scarpetta$^{a,c}$  \\
{\em $^a$ Dipartimento di Scienze Fisiche ``E.R. Caianiello''}\\ 
{\em Universit\`a di Salerno, 84081 Baronissi (Salerno), Italy.}\\
{\em $^a$ Istituto Nazionale di Fisica Nucleare, Sezione di Napoli,}\\
{\em $^b$ Department of Physics, University of Regina,} \\
{\em Regina, Sask. S4S 0A2, Canada.} \\
{\em $^c$ International Institute for Advanced Scientific Studies} \\
{\em Vietri sul Mare (SA), Italy.} \\
}
              \date{\empty}
              \maketitle

\begin{abstract}

Caianiello's quantum geometrical model with maximal acceleration places the upper limit 
$\mu\leq 719.5$ GeV on the mass of the Higgs boson. The model also provides
an equation linking the mass of the W boson to the Higgs mass
$\mu$ and independent symmetry breaking and  
mass generating mechanisms. These may further restrict the value of $\mu$.

\end{abstract}  
                               
\thispagestyle{empty}          
\vspace{20. mm}                
\vspace{5. mm}                 
              \vfill           
	      \end{titlepage}  

\section{Introduction}

In a recent work Kuwata \cite{KUW}  points out that the existence of a 
maximal acceleration (MA), conjectured on different grounds by several authors 
\cite{981}, \cite{CROSS},\cite{STAR}
would set an upper limit on the mass of the Higgs 
boson. Kuwata argues that the onset of MA would transform the 
Higgs--fermion interaction potential, 
assumed to be of the Yukawa type, into a low order polynomial
in $r$ at distances less than a critical value $r_c$
. By restricting the polynomial to third order, which requires the MA
to occur at $r=0$, and by demanding the continuity of the two potentials at
$r=r_c$ up to second order derivatives, the short range potential can be 
fully determined. The Higgs Lagrangian is then modified to accomodate 
the new potential. Standard model relations lead to the upper limit 
$\mu=M_W\sqrt{1.77 k \sin^2\theta_W/\alpha}$, where $M_W$ is the mass of the 
$W$-boson, $\theta_W$ is the Weinberg angle and $\alpha$ the fine structure 
constant. $k\simeq 1$ is a factor that appears in the definition of the
maximal acceleration ${\cal A}=km_r$ $(\hbar=c=1)$ adopted 
by Kuwata and $m_r$ is the reduced fermion mass.

The expression for ${\cal A}_m$ used by Kuwata was originally derived by Caianiello 
\cite{981}, \cite{984}, and subsequently re-derived by other authors \cite{GP}, by using quantum 
mechanics and Landau's theory
of fluctuations. It does not require additional assumptions and is, as such, a concept in search of a deeper interpretation. In related work, in fact, Caianiello and collaborators provided a geometrical 
framework in which the MA of a particle directly affects the spacetime experienced 
by the particle itself \cite{CCROSS}, \cite{IMM}. According to this model, a particle of mass $m$ 
accelerating along its worldline experiences a spacetime of metric 
\begin{equation}\label{eq1}
d\tilde{s}^2=ds^2\left(1+\frac{\eta_{\mu\nu}\ddot{x^{\mu}}
\ddot{x^{\nu}}}{{\cal A}_m^2}
\right)=\sigma^2(x)ds^2\, {,}
\end{equation}
where $ds^2$ refers to Minkowski space whose metric $\eta_{\mu\nu}$ has
signature $-2$ and $\ddot{x}^{\mu}\equiv d^2x^{\mu}/ds^2$ is 
the four--acceleration of the particle \cite{981}.  
Unless explicitly stated, we choose $k=2$ \cite{GP}.

Generalizations to include background gravitational fields can be obtained by
replacing $\eta_{\mu\nu}$ with the corresponding metric tensors. 
Several important implications of Eq.(\ref{eq1}) have already been discussed 
in the literature \cite{LAM}, \cite{LET},\cite{NS}, \cite{REL}. 
The purpose of this paper is to show that Eq.(\ref{eq1})
also places limits on the mass of the Higgs boson. These limits do not require additional changes to the Higgs Lagrangian, are consistent with those  of 
Ref.\cite{KUW}, when allowance is made for the different values of $k$, 
and with the experimental limits $50$GeV$\leq\mu\leq1000$GeV obtained
 by using radiative corrections \cite{EXC}. They are derived and discussed 
in Section 2. Section 3 contains a brief derivation of the Dirac equation 
for the metric (\ref{eq1})
and a discussion of its most noticeable consequences. Eq. (\ref{eq1})
also allows the derivation of a formula relating the masses of the
standard model fermions and bosons to that of the Higgs boson.
The relation between MA and standard model is further discussed
in Section 5 where a mass generating mechanism is also introduced.
The conclusions are presented in Section 6. 

\section{Limits to the Higgs mass}

It is possible to establish an upper limit to $\mu$ 
starting from Eq. (\ref{eq1}) directly. An explicit expression for
$\sigma(x)$ can be obtained classically by assuming, as in Ref.\cite{KUW}, that 
the fermion--fermion interaction can be represented by a Yukawa potential
$$
V(r)=\frac{g_H^2}{4\pi}\,\frac{e^{-\mu r}}{r}\,{.}
$$
One immediately finds
\begin{equation}\label{eq2}
\sigma(r)=\sqrt{1-\left(\frac{1}{\pi v^2}\right)^2\frac{e^{-2\mu r}(1+
\mu r)^2}{k^2r^4}}\, {,}
\end{equation}
where $v=\sqrt{2}<0\vert\phi\vert0>=246.2185$GeV is the usual 
standard model parameter.

Since $d\tilde{s}$ is real, $\sigma(r)$ must remain real for all values of
$\mu$ and $r$. When $\vert\ddot{x}\vert={\cal A}_m$, $\sigma(r)$ vanishes. 
On the other hand $\sigma(r)\sim 1$ for $r>>1/\mu$. At these distances 
the fermions do not
interact appreciably with the Higgs boson. A function $r_c(\mu)$ must therefore
exist such that $\sigma(r)\geq 0$ for $r\geq r_c(\mu)$. An expression for 
$r_c(\mu)$ can be  obtained by expanding $\sigma(r)$ to third order in
the neighborhood of $1/\mu$, where the
acceleration presumably reaches its highest values, and by 
equating the result to zero. The expansion is given by
\begin{equation}\label{eq3}
\sigma(r)\sim B + \frac{9.32758\cdot 10^{-12}\mu^5}{B}(r-\frac{1}{\mu})+
\end{equation}
$$
\frac{1}{2}\left(-\frac{8.70037\cdot 10^{-23}\mu^{10}}{B^3}-
\frac{5.31672\cdot 10^{-11}\mu^6}{B^7}\right)\left(r-\frac{1}{\mu}\right)^2+
$$
$$
+\frac{1}{3}\left[\frac{1.7256\cdot 10^{-17}\mu^7}{B}+
  \right.
$$
$$
\left. 
+\frac{1.39914\cdot 10^{-11}\mu^5}{B}\left(\frac{8.70037\cdot 10^{-23}\mu^{10}}{B^4}
+\frac{5.31672\cdot 10^{-11}\mu^6}{B^2}\right)\right]
\left(r-\frac{1}{\mu}\right)^3\,{,}
$$
where $B=(1-3.73103\cdot 10^{-12}\mu^4)^{1/2}$. Expression (\ref{eq3})
is real if $B$ is real, which yields $\mu\le 719.52$GeV. Of the three solutions
for $r$ obtained by equating (\ref{eq3}) to zero, only one is real. 

It  also follows that
$r\ge 0$ for $316.1$ GeV$\leq\mu\leq719.5$ GeV. These are the limits of validity 
of the solution. They are not affected greatly by adding additional terms to the
expansion for $\sigma(r)$. At order $(r-1/\mu)^5$ one finds in fact $225.4$ GeV
$\leq\mu\leq 719.53$ GeV for $k=2$.

In this model it seems therefore possible to establish a lower
limit in addition to an upper one. However the lower limit is only a construct
of the approximation, devoid of physical meaning. Eq (\ref{eq2}) 
does in fact diverge at $r=0$.
For each value of $\mu$ there is a value 
of $r$ for which $\sigma=0$. For the choice $k=1$ of Kuwata, the corresponding
limits at $O((r-1/\mu)^3)$ are $223.5$ GeV$\leq\mu\leq 508.8$ GeV. 
In both instances the upper limits are well above 
the energy ranges of recent searches \cite{ALE}.
Contrary to Kuwata's results, the MA is reached for $r=0$ only for $\mu$ at the
lower limit, where the solution becomes inaccurate. 
Eq. (\ref{eq1}) also affects the fermion--fermion interaction.
This is shown in the next section.

\section{The covariant Dirac equation}

With the introduction of the metric tensor 
$\tilde{g}_{\mu\nu}=\sigma^2(x){\eta}_{\mu\nu}$, the Dirac equation must be
generalized to curved space--time and connected to a local Minkowski frame
by means of the vierbein field $e_{\mu}^a(x)$, where Latin indices refer to 
the local frame and Greek indices to a generic non--inertial frame. The
vierbeins follow from (\ref{eq1}) and are given by $e_{\mu}^a(x)=\sigma(x)
\delta_{\mu}^a$. The covariant Dirac equation has the form
\begin{equation}\label{eq3.1}
[i\gamma^{\mu}(x){\cal D}_{\mu}-V(r)-m]\psi(x)=0\,{.}
\end{equation}
The matrices $\gamma^{\mu}(x)$ satisfy the anticommutation relations
$\{\gamma^{\mu}(x),\gamma^{\nu}(x)\}=2\tilde{g}^{\mu\nu}(x)$. The covariant
derivative ${\cal D}_{\mu}\equiv \partial_{\mu}+\omega_{\mu}(x)$
contains the total connection $\omega_{\mu}=\sigma^{ab}\omega_{\mu ab}$,
where $\sigma^{ab}=[\gamma^a, \gamma^b]/4$,
$\omega_{\mu\,\,\,\,b}^{\,\,\,\,a}=
(\Gamma^{\lambda}_{\mu\nu}e_{\lambda}^{\,\,\,\,a}-
\partial_{\mu}e_{\nu}^{\,\,\,\,a})e^{\nu}_{\,\,\,\,b}$ and 
$\Gamma^{\lambda}_{\mu\nu}$ represent the usual Christoffel symbols.
For conformally flat metrics $\omega_{\mu}=\sigma^{ab}\eta_{a\mu}
(\ln\sigma)_{,b}$. By using the transformations $\gamma^{\mu}(x)=e^{\mu}_a
\gamma^a=\sigma^{-1}(x)\gamma^{\mu}$, where $\gamma^{\mu}$ are the
usual position--independent Dirac matrices, Eq. (\ref{eq3.1}) becomes
\begin{equation}\label{eq3.2}
\left[i\gamma^{\mu}\partial_{\mu}+i\frac{3}{2}\gamma^{\mu}(\ln\sigma)_{,\mu}
-m\sigma-V(r)\sigma\right]\psi(x)=0\,{.}
\end{equation}
\noindent From Eq. (\ref{eq3.2}) one finds the Hamiltonian
\begin{equation}\label{eq3.3}
H=\vec{\alpha}\cdot\vec{p}+\gamma^0V(r)\sigma(r)-
i\frac{3}{2}\gamma^0\gamma^{\mu}(\ln\sigma)_{,\mu}+m\sigma(r)\gamma^0\,{.}
\end{equation}
The line element (\ref{eq1}) therefore induces additional potential terms in 
the Hamiltonian. They are represented by the last two terms in Eq. 
(\ref{eq3.3}) and give rise to a combined potential which is in general
spin--dependent and highly repulsive for $\sigma(r)< 1$.

In the present calculation $\sigma(r)$, given by Eq. (\ref{eq2}), is 
time--independent, hence all potential terms are conservative \cite{PAR}. Classically
the fermions will therefore accelerate toward the centre of mass while
adjusting their speeds to the height of the potential barrier they move over.
Once they reach the barrier, they must bounce back.
Numerical studies of these terms indicate that when $\mu\simeq 700$ GeV and
the total fermion energy is also close to $700$ GeV, the bounce back occurs at 
$r\simeq 0.00184$ GeV$^{-1}$, while $\sigma$ vanishes at 
$r\simeq 0.0013975$ GeV$^{-1}$. The highest value of the acceleration reached 
by the fermions at $r\simeq 0.00184$ GeV$^{-1}$ is $\sim 0.49 {\cal A}_m$. For 
$\mu\sim 500$ GeV and a total fermion energy also of $500$ GeV the barrier is
hit at $r\simeq 0.002037$ GeV$^{-1}$ where the acceleration is 
$\sim 0.46 {\cal A}_m$.
Only at the lower limit $\mu=316.1$ GeV$^{-1}$ is $\vert\ddot{x}\vert={\cal A}_m$ 
at $r=0$.
With the exception of this limiting case, a consequence of the approximations made, 
Kuwata's assumption that the MA is 
reached at $r=0$ during the interaction cannot be made in the present model. It also follows that the limits determined in Sect. 2, 
corresponding to accelerations undamped by the repulsive barrier, represent
indeed the extreme upper limits derivable from this geometrical
model.

Qualitatively one may also conclude that as the two fermions approach 
the limiting value of the potential and their relative speed becomes non
relativistic, the term $\gamma^0\gamma^{\mu}(\ln\sigma)_{,\mu}$, that
connects the small spinor components, is switched off. The residual potential
then becomes slightly negative allowing the fermions to weakly bind to
each other already for $\mu\sim 400$ GeV. No quantitative analysis of the effect of the repulsive 
barrier has so far been
carried out for the quantum scattering problem. 
The barrier effect on the energy spectrum has however been 
analyzed in detail in the case of hydrogenic atoms \cite{LAM}.

A study of the curvature invariant
\begin{equation}\label{eq3.4}
R=-\frac{6}{r}\frac{1}{\sigma^3(r)}[2\sigma^{\prime}(r)+
r\sigma^{\prime\prime}(r)]
\end{equation}
indicates that R is always positive or null and grows rapidly in the neighborhood 
of those
values of $r$ and $\mu$ for which $\sigma =0$, as expected.

\section{Scale changes}

It is interesting to observe that for values of $g_H$ given by the standard
model, $\sigma(r)$, represented by (\ref{eq2}), is independent of the mass of the
accelerated fermions and depends only on the mass of the Higgs boson.
The same applies to the accelerated $W$ and $Z$ bosons in the Yukawa potential
approximation of Sect. 2.
It now follows from Eq. (\ref{eq1}) that if the 
characteristic length of an accelerated particle is  
$ds\approx m^{-1}$, $ds$ will contract to a length 
$\tilde{ds}\approx \mu^{-1}$ when the particles are
accelerated. The contraction is due to the conformal factor $\sigma$ which provides an immediate, approximate relationship between $m$ and $\mu$.
From Eq. (\ref{eq1}) and (\ref{eq2}) one obtains the formula

\begin{equation}\label{eq8}
m=\mu\sqrt{1-\frac{4\mu^4}{(\pi ev^2)^2k^2}}\, {,}
\end{equation}
where $(\pi ev^2)^{-2}\sim 3.73103\cdot 10^{-12}$GeV$^{-4}$ and $e$ is
Neper's number. Given $\mu$, one may in principle derive the values of $m$ from Eq. (\ref{eq8}). In fact this equation has always the real and positive solution 
$$
\mu\simeq m\quad\quad \mbox{for} \quad\quad 
\frac{4m^4}{(\pi ev^2)^2k^2}\ll1\,{,}
$$
corresponding to the uncontracted length, and the solution
$$
\mu\simeq \sqrt{\frac{\pi ev^2 k}{2}}\quad\quad \mbox{for} 
\quad\quad m^2\sqrt{\frac{2}{\pi ev^2 k}}\ll 1\,{.}
$$ 
 The latter solution gives the value of $\mu$ at which a change in scale occurs. On applying (\ref{eq8}) to scattering processes involving the
fermions and bosons of the standard model 
one obtains the solutions given in Table I.
The results for the lower fermion masses are very sensitive to the value of 
$\mu$. On using the formula \cite{UNK}
$$
m_t=180\pm7+13\ln\frac{\mu}{300\mbox{GeV}}\,{,}
$$
and the values of $\mu$ determined from Eq. (\ref{eq8}) one finds for $k=2$
$$
 m_t=180\pm7+11.372=\left\{ \begin{array}{c} 
                                 198.372\mbox{GeV} \\
                                 184.372\mbox{GeV}
                              \end{array}
                       \right.  
$$
and for $k=1$
$$
 m_t=180\pm7+6.867=\left\{ \begin{array}{c} 
                                 193.867\mbox{GeV} \\
                                 178.867\mbox{GeV.}
                              \end{array}
                       \right.  
$$
All these values are consistent with the global fit to all data given in 
Ref.\cite{UNK}
$$
m_t=180\pm 7^{+12}_{-13}\mbox{GeV}
$$
and with the values of $m_t$ obtained from (\ref{eq8}).
Finally, the solution $\mu\simeq \sqrt{\pi e v^2 k/2}$ may be written as
\begin{equation}\label{eq8b}
\mu=M_W\frac{2}{g}\sqrt{\frac{\pi ek}{2}}=
M_W\sqrt{\frac{1.36k\sin^2\theta_W}{\alpha}}
\end{equation}
in good agreement with Kuwata's formula given in Sect. 1 and the limit of Sect. 2.

\section{\bf MA and the Standard Model}

The
behaviour of the Dirac equation at $\sigma=0$ is pathological. 
The particle's behaviour is however well defined for increasing acceleration up
to, but excluding, the MA. For decreasing $\sigma(r)$ the particle's effective mass,
$m\sigma(r)$, decreases and so does the term $V(r)\sigma(r)$. 
These features aptly embody the notion of asymptotic freedom in the
model. On the other hand, the repulsive barrier grows logarithmically, which
indicates that the MA is classically unattainable. 

Eq. (\ref{eq3.2}) remains invariant in form under the 
transformation $\sigma\to
i\vert\sigma\vert $. Fermions with acceleration higher than the MA
may therefore be thought of as particles obeying the Dirac equation,
but in regions of space--time where the roles of space and time
are interchanged relative to the usual ones.

The two regions, $\sigma$ imaginary (I) and $\sigma$ real (II) are  
disconnected.
The detailed behaviour in the neighborhood of $\sigma =0$ will in general 
depend on $g_{\mu\nu}$ in the metric $\tilde{g}_{\mu\nu}=
\sigma^2g_{\mu\nu}$. 
In I the light cone is rotated by $\pi/2$ relative to II. Fermions may possibly travel
from I to II and II to I but only as far as $r_c$ ($\sigma=0$). 
The fermions are confined 
to their world tubes and that of the Higgs boson by the action of $V(r)\sigma(r)$ but are prevented from
reaching the MA by the combined action of $m\sigma(r)$ and the logarithmic
potential.

The space--time experienced by the accelerating particle has, however, nonvanishing 
curvature inside the world tube. This leads to interesting consequences. 
A Lagrangian term $(\xi/2)R(x)\sigma^2(x)\vert\phi(x)\vert^2$, 
where $\xi$ is a
parameter, usually accompanies any field capable of interacting with
gravity. Though $R$ is really experienced only by the accelerating particle,
its dynamics in the field that produces acceleration must be 
effectively consistent with the presence of
this interaction term. For Higgs fields one must have $\xi\leq 0$ or
$\xi\geq 1/6$ \cite{HOS}. The Higgs potential then becomes
\begin{equation}\label{Hpot}
{\it V}(\phi)=\left(\bar{\mu}^2\vert\phi\vert^2 + \frac{1}{2}\xi 
R\vert\phi\vert^2 + \lambda\vert\phi\vert^4\right)
\sigma^2\,{,}
\end{equation}
where $\bar{\mu}$ is the usual mass parameter that appears in the Higgs mechanism
and $\lambda >0$. Formally, the vacuum has minima at
\begin{equation}\label{min}
\vert\phi\vert=\pm\sqrt{-\frac{\bar{\mu}^2+
(\xi/2)R}{2\lambda}}\equiv \frac{v}{\sqrt{2}}
\end{equation}
and the mass of the Higgs boson becomes
\begin{equation}\label{Hmass}
\mu^2=-2\left(\bar{\mu}^2+\frac{\xi}{2}R\right)\sigma^2=2\lambda v^2\sigma^2\,{,}
\end{equation} 
where, as usual, $\mu^2$=2$\lambda v^2$ for $\sigma^2= 1$.
It follows from (\ref{min}) and (\ref{Hmass}) that the MA provides an additional
independent symmetry breaking mechanism which is present even when $\bar{\mu}=0$.

Several possibilities now exist: \\
i) $\xi>0$. If in addition $\bar{\mu}^2>0$,
then $\vert\phi\vert$ in Eq. (\ref{min}) becomes imaginary, the only minimum 
of ${\it V}(\phi)$ is at $\vert\phi\vert=0$ and the symmetry is not broken. 
If, on the contrary, $\bar{\mu}^2<0$ as required by the usual Higgs mechanism, then
one can have either $\vert\bar{\mu}\vert^2>(\xi/2)R$, which leads to spontaneous 
symmetry breaking and minima at 
$$
\vert\phi\vert =\pm\sqrt{\frac{\vert\bar{\mu}\vert^2-(\xi/2)R}{2\lambda}}\,{,}
$$
or $\vert\bar{\mu}\vert^2<(\xi/2)R$ with the minimum at $\vert\phi\vert =0$
as before. The particular value $\vert\bar{\mu}\vert^2=(\xi/2)R$ restores the
original unbroken symmetry and implies $\mu =0$. \\
ii) $\xi<0$. If one also has $\bar{\mu}^2>0$, two additional cases are possible:
a) $\bar{\mu}^2>(\vert\xi\vert/2)R$ . The only minimum of Eq. (\ref{Hpot})
is again at $\vert\phi\vert =0$; b) $\bar{\mu}^2<(\vert\xi\vert/2)R$. The
new minima occur at 
$$
\vert\phi\vert=\pm\sqrt{\frac{\vert\bar{\mu}^2-(\vert\xi\vert/2)R\vert}{2\lambda}}
$$
and the symmetry is broken.  Finally, if $\bar{\mu}^2<0$, the symmetry is again
broken and the vacuum minima are at
$$
\vert\phi\vert=\pm\sqrt{\frac{\vert\bar{\mu}^2\vert+(\vert\xi\vert/2)R}{2\lambda}}
\,{.}
$$
The particular value $\bar{\mu}^2=(\vert\xi\vert/2)R$ also restores the original
vacuum value $\vert\phi\vert =0$. In all instances the corresponding values of 
$\mu$ can be easily derived from Eq. (\ref{Hmass}).

All possibilities require some sort of mass generating mechanism to be associated 
with $\xi R\sigma^2$. In order to show that this mechanism is present, the
choice $\bar{\mu}=0$ in Eq. (\ref{min}) is particularly appropriate. On averaging
$\sigma^2 R$ over three dimensional space 
\begin{equation}\label{aver}
<\sigma^2 R>=\frac{\int_0^{1/\mu}\sigma^6(r)R(r)r^2dr}{\int_0^{1/\mu}
\sigma^4(r)r^2dr}
\end{equation}
and expanding the integrands to order $(r-1/\mu)^5$, one finds that
the equation
\begin{equation}\label{avereq}
\vert\xi\vert<\sigma^2 R> - \mu^2 = 0
\end{equation}
has solutions for $\vert\xi\vert\geq 0$. For $(1/100)\leq\vert\xi\vert\leq 100$ one 
gets $420$ GeV$\leq\mu\leq 635$ GeV. The value $\mu\sim 555$ GeV occurs at 
$\vert\xi\vert =1/6$. For $\vert\xi\vert =1$ one finds $\mu\sim 615$ GeV and
$\mu$ does not increase beyond $635$ GeV for $\vert\xi\vert\geq 100$.
In fact no value of $\vert\xi\vert$ exists that can move $\mu$ to the
upper limit of Sect. 2. The case $k=1$ is entirely similar.
For $(1/100)\leq\xi\leq 100$ one finds $296$ GeV $\leq \mu\leq 450$ GeV with
$\mu \sim 392$ GeV at $\vert\xi\vert =1/6$ and increasing from 
$435$ GeV at $\vert\xi\vert =1$ to $450$ GeV for higher value of
$\vert\xi\vert$. Here too no value of $\vert\xi\vert$ yields 
$\mu\sim 508$ GeV, the upper limit discussed in Sect. 2.

\section{Conclusions}

Kuwata's derivation of an upper limit for the value of the Higgs boson mass is based
on the maximal value that non--relativistic quantum mechanics places on the acceleration
of a particle. It is, as such, perfectly legitimate.

Caianiello's geometrical model, on the other hand, is more complete. It incorporates
the notion of MA, but also contains detailed dynamical prescriptions for the 
behaviour of an accelerating particle. These are essentially contained in Eq. 
(\ref{eq1}) and its effect on the actual form of the Dirac equation. The reality of
$\sigma(r)$ already is sufficient to place an upper limit on the mass of the Higgs 
boson. For the sake of comparison with Kuwata's results estimates have been given
for both $k=2$ and $k=1$ (Kuwata's choice), though present measurements of the Lamb 
shift in hydrogenic  atoms agree with $k=2$, but not with $k=1$.

The result, $\mu\leq 719.5$ GeV, must indeed be considered as an upper limit 
because the actual value of the MA is not reached by the particle, 
as derived in Sect. 3 from the covariant Dirac equation.
In a non-geometrical approach the Dirac equation takes its usual Minkowski space form and no information is available about the additional potential terms present in Eq. (\ref{eq3.3}).
Using Eq. (\ref{eq1}) one can also derive Eq. (\ref{eq8}) which links the
masses of all particles in the standard model to the mass of the Higgs boson.
This equation reproduces all values of fermions and bosons well (Table I): they correspond to their original uncontracted scales. A change in scale occurs however for the solution Eq. (\ref{eq8b}) that agrees well, for $k=1$, with  Kuwata's result and the limit of Sect. 2. Section
5 discusses a number of implications that (\ref{eq1}) has for the standard model.
In general one may conclude that (\ref{eq1}) provides an additional
independent symmetry breaking mechanism that may affect the vacuum minima
in a variety of ways. A mass generating mechanism can be associated 
with the average curvature experienced by an accelerating particle.
When the mass parameter $\bar{\mu}$ of the standard model vanishes,
Eq. (\ref{Hmass}) for the Higgs boson mass has solutions 
$420$ GeV $\leq \mu\leq 635$ GeV over a wide range of values of the
parameter $\xi$. The additional restriction introduced by the mass generation 
mechanism is perfectly compatible with the upper limit $\mu\leq 719.52$ GeV
imposed by the reality of the conformal factor $\sigma (r)$.

\bigskip
\begin{centerline}
{\bf Acknowledgments}
\end{centerline}

Research supported by NATO Collaborative Research Grant No. 970150, 
Ministero dell'Universit\`a e della Ricerca Scientifica of Italy
and the Natural Sciences and
Engineering Research Council of Canada. 

G.P. gladly acknowledges the continued research support 
of Dr. K. Denford, Dean of Science, University of Regina.

G.L. and G.S. wish to thank Dean Denford for his kind hospitality 
during stays at the University of Regina.
 
\newpage

\begin{center}
Table I  
\end{center}
\begin{center}
\begin{tabular}{c|c|c||c|c}\hline
            &   $k=2$         &   $k=2$            &  $k=1$          &  $k=1$ \\
Interaction & $m(\mbox{GeV})$ &  $\mu(\mbox{GeV})$ & $m(\mbox{GeV})$ & 
                                                             $\mu(\mbox{GeV})$\\
                                                                       \hline
$Hee$  & $0.511\cdot 10^{-3}$ & 719.52  & $0.511\cdot 10^{-3}$ & 508.8 \\
$Huu$  & $5\cdot 10^{-3}$ & 719.52  & $5\cdot 10^{-3}$ & 508.8 \\
$Hdd$  & $10^{-2}$ & 719.52  & $10^{-2}$ & 508.8 \\ 
$Hbb$  & 4.3       & 719.51  & 4.3       & 508.8 \\ 
$Htt$  & 180.716   & 707.54   & 181.8    & 490.7 \\ 
$Hss$  & 0.34      & 719.52   & 0.34     & 508.8 \\ 
$Hcc$  & 1.3       & 719.52   & 1.3      & 508.8 \\ 
$HWW$ & 80.33    & 717.25   & 80.4     & 505.5  \\ 
$HZZ$ & 91.2     & 716.6    & 91.2     & 504.6   \\ \hline
\end{tabular}
\end{center}

\vfill

\end{document}